\documentclass[12pt]{report}
\usepackage{graphicx,amsfonts,amsmath,amssymb,mathrsfs}

\title{Bohmian Mechanics}
\author{
   Detlef D\"urr\footnote{Mathematisches Institut, Ludwig-Maximilians-Universit\"{a}t,
         Theresienstra{\ss}e 39, 80333 M\"{u}nchen, Germany.
         E-mail: duerr@mathematik.uni-muenchen.de},
   Sheldon Goldstein\footnote{Departments of Mathematics and Physics,
         Rutgers University, Hill Center, 
         110 Frelinghuysen Road, Piscataway, NJ 08854-8019, USA.
         E-mail: oldstein@math.rutgers.edu},\\
   Roderich Tumulka\footnote{Department of Mathematics,
         Rutgers University, Hill Center, 
         110 Frelinghuysen Road, Piscataway, NJ 08854-8019, USA.
         E-mail: tumulka@math.rutgers.edu}, and
   Nino Zangh\`\i\footnote{Dipartimento di Fisica and INFN sezione di
         Genova, Via Dodecaneso 33, 16146 Genova, Italy. E-mail: 
         zanghi@ge.infn.it}
}
\date{January 6, 2008}

\newcommand{\Hilbert}{\mathscr{H}}

\newcommand{\RRR}{\mathbb{R}}
\newcommand{\CCC}{\mathbb{C}}
\newcommand{\scp}[2]{\langle #1|#2 \rangle}

\newcommand{\vQ}{\boldsymbol{Q}}

\newcommand{\im}{\mathrm{Im}}
\newcommand{\be}{\begin{equation}}
\newcommand{\ee}{\end{equation}}
\newcommand{\see}{}%{$\to$}

\begin{document}
\maketitle

\noindent \textbf{Bohmian mechanics} is a theory about point  particles moving along trajectories. It has the property that in a world governed by Bohmian mechanics, observers see the same statistics for experimental results as predicted by quantum mechanics. Bohmian mechanics thus provides an \emph{explanation} of quantum mechanics.  Moreover, the Bohmian trajectories are defined in a non-conspiratorial way by a few simple laws.

\bigskip

\noindent\textbf{Overview.}
          Bohmian mechanics is a version of quantum mechanics for
          nonrelativistic particles in which the word ``particle'' is to be
          understood literally: In Bohmian mechanics quantum particles have
          positions, always, and follow trajectories.
These trajectories differ, however, from the classical Newtonian trajectories. Indeed, the law of motion, see eq.~\eqref{Bohm} below, involves a wave function. As a consequence, the role of the wave function in Bohmian mechanics is to \emph{tell the matter how to move}. 

Bohmian mechanics constitutes a \emph{quantum theory without observers}, i.e., a theory that is formulated not in terms of what observers see but in terms of objective events, regardless of whether or not they are observed. Bohmian mechanics provides a consistent resolution of all paradoxes of quantum mechanics, in particular of the so-called \emph{measurement problem}. In particular, the \emph{collapse of the wave function} can be derived from Bohmian mechanics.

Bohmian mechanics is sometimes called a \emph{hidden variables theory} because it involves variables besides the wave function. However, there is a danger of confusion here because the term ``hidden variables theory'' is often used to convey the idea that every ``quantum measurement'' of an ``observable'' reveals a pre-existing value of that observable, which is not the case in Bohmian mechanics.

Bohmian mechanics is \emph{deterministic}.  But the motivation behind Bohmian mechanics is not to obtain a deterministic theory, but rather to obtain a coherent account of the nature of physical reality. In this regard, we note that some variants of Bohmian mechanics, developed by its proponents, are stochastic rather than deterministic, for example Bell's proposal for lattice quantum field theory \cite{Bell86}. 

Historically, the ``Bohmian'' law of motion, see eq.~\eqref{Bohm} below,
was first proposed by de Broglie \cite{deB}. However, Bohm \cite{Bohm52}
was the first to recognize that this theory explains all of the phenomena
of (non-relativistic) quantum mechanics. %On the other hand,  and Rosen \cite{Rosen}, and similar equations were proposed even earlier by Slater. 

\bigskip
%\section{Postulates and Consequences}

\noindent\textbf{Defining Equations.}
Bohmian mechanics is a non-relativistic theory governing the behavior of a system of $N$ point particles moving in physical space $\RRR^3$ along trajectories. Let $\vQ_i(t)\in\RRR^3$ denote the position of the $i$-th particle of the system at time $t$, and $Q(t) = \bigl(\vQ_1(t),\ldots, \vQ_N(t)\bigr)\in\RRR^{3N}$ its configuration. 

The trajectories are governed by \emph{Bohm's law of motion} \cite{Bohm52,Bell66}
\begin{equation}\label{Bohm}
\frac{d\vQ_i}{dt} = \frac{\hbar}{m_i} \im \frac{\Psi_t^* \nabla_i \Psi_t}{\Psi_t^* \Psi_t}\bigl(Q(t)\bigr)\,,
\end{equation}
where $m_i$ is the mass of particle $i$, $\im$ denotes the imaginary part,
$\Psi_t:\RRR^{3N} \to \CCC^k$ (i.e., a function of the configuration with $k$ complex components) is the wave function at time $t$, $\Phi^* \Psi$ is the scalar product in $\CCC^k$, and $\nabla_i$ is the gradient relative to the 3 coordinates of particle $i$. (In case $k=1$, i.e., for complex-valued wave functions, a factor $\Psi_t^*$ cancels on the right hand side of \eqref{Bohm}.)

The wave function evolves according to the \emph{Schr\"odinger equation}
\begin{equation}\label{Schr}
i\hbar \frac{\partial \Psi_t}{\partial t} = -\sum_{i=1}^N \frac{\hbar^2}{2m_i} \nabla_i^2 \Psi_t + V \Psi_t\,, 
\end{equation}
where $V: \RRR^{3N}\to\RRR$ is the potential function. (The potential, while often assumed to be real-valued, may take values in the space of self-adjoint complex $k\times k$ matrices instead of $\RRR$.) The wave function is postulated to belong to the \see Hilbert space $\Hilbert = L^2(\RRR^{3N},\CCC^k)$ of square-integrable functions (and to be sufficiently smooth).

%A third postulate (which may or may not be included in the definition of Bohmian mechanics) asserts that on the universal level \emph{the initial configuration $Q(t_0)$ is typical relative to the $|\Psi_{t_0}|^2$ distribution}. We will elucidate later what that means.

\bigskip

\noindent\textbf{Deterministic Evolution.}
Since the Schr\"odinger equation does not involve the particle positions $\vQ_i(t)$, it can be solved first and determines the wave function $\Psi_t$ for every time $t$ once an initial wave function $\Psi_{t_0}$ is specified for any time $t_0$ that we choose to regard as the initial time. Next note that the right hand side of \eqref{Bohm} consists of the  3 components corresponding to particle $i$ out of the $3N$ components of a vector field $v^{\Psi_t}$ on configuration space $\RRR^{3N}$. As a consequence, the equations \eqref{Bohm} for all $i=1,\ldots,N$ can be summarized by
\begin{equation}\label{v}
\frac{dQ}{dt} = v^{\Psi_t}\bigl(Q(t)\bigr)\,.
\end{equation}
Regarding $\Psi_t$ as known, this is a (time-dependent) ordinary differential equation (ODE) of first order, and as such determines the entire history $t\mapsto Q(t)$ once an initial configuration $Q(t_0)$ is specified. That is why Bohmian mechanics is \emph{deterministic}: once $Q(t_0)$ and $\Psi_{t_0}$ are specified, the entire history is fixed by the equations \eqref{Bohm} and \eqref{Schr}. This fact also implies that the pair $(Q(t_0), \Psi_{t_0})$ can be regarded as the \emph{state} of the Bohmian particle system at time $t_0$. Since the choice of $t_0$ is arbitrary, the state at any time $t$ is the pair $(Q(t),\Psi_t)$, and the phase space of Bohmian mechanics is $\RRR^{3N} \times \Hilbert$.

\bigskip

\noindent\textbf{System or Universe.}
The equations of Bohmian mechanics could be applied to a familiar system (e.g., an atom) or to the universe as a whole. Of course, one cannot expect that the equations hold for every system, for example for systems that interact with their environments. So let us begin with the system for which the equations are primarily intended: the universe. In this setting, $N$ is the number of particles in the universe, and $\Psi_t$ is the wave function of the universe. To consider such a wave function is unusual; after all, the quantum formalism never refers to a wave function of the universe; the quantum formalism, providing the probabilities for the results of observations performed on a system by an external observer, involves the wave function of that system and not of the entire universe. In the context of Bohmian mechanics, however, the wave function of the universe is not at all a meaningless concept, as it influences the motion of the particles according to \eqref{Bohm}. 

When \eqref{Bohm} and \eqref{Schr} hold for the universe, it follows that equations of the same type (but with smaller $N$) hold for certain subsystems. 
(We shall assume here for simplicity that $k=1$, i.e., that we are dealing with spinless particles.) Consider a subsystem of the universe with configuration $X$ (the $x$-system), so that the configuration $Q$ of the universe is of the form $Q=(X,Y)$ with $Y$ the configuration of the \emph{environment} of the $x$-system. Then a natural notion of the wave function of the $x$-system is provided by its \emph{conditional wave function}
\be\label{condwf}
\psi(x) = \Psi(x,Y)\,,
\ee
where $\Psi(q) = \Psi(x,y)$ is the wave function of the universe. It is easy to see that the $x$-system obeys \eqref{v} (with $Q=X$ and $\Psi=\psi$).

Moreover, if the $x$-system is suitably decoupled from its environment, \eqref{Schr} will hold as well. 
For example, this is the case when there is no interaction between the
$x$-system and its environment, and the wave function of the universe is of
the form
\begin{equation}\label{effectivewf}
\Psi(x,y) = \psi(x)\, \varphi(y) + \Phi(x,y)
\end{equation}
with $\varphi$ and $\Phi$ having macroscopically disjoint $y$-supports (so
that they will never again overlap appreciably), and with $Y$ lying in the
support of $\varphi$. Such a situation often arises after a ``quantum
measurement.''
%More generally, even when \eqref{effectivewf} does not hold, Bohmian mechanics provides a notion of a wave function $\psi$ of a subsystem: the \emph{conditional wave function}
%\begin{equation}\label{condwf}
%\psi(x) = \mathcal{N} \, \Psi(x,Y)\,,
%\end{equation}
%where $\mathcal{N}$ is the normalizing factor, $\Psi(x,y)$ the wave function of the universe, and $Y$ the (actual) configuration of the system's environment.

\bigskip

\noindent\textbf{Equivariance.}
If the initial configuration $Q(t_0)$ is chosen at random with probability density $|\Psi_{t_0}|^2$ then the configuration $Q(t)$ at any other time $t$ is random with probability density $|\Psi_t|^2$. (Whenever speaking of probabilities, we assume that $\Psi$ has been normalized, by multiplication by a suitable constant, so that $\scp{\Psi}{\Psi}=\int |\Psi(q)|^2 dq=1$.) This fact, known as \emph{equivariance}, follows from the continuity equation
\begin{equation}\label{continuity}
\frac{\partial\rho}{\partial t} = - \nabla \cdot (\rho \, v)
\end{equation}
for $\rho = |\Psi|^2$ and with the Bohmian velocity vector field $v=v^\Psi$ as in \eqref{v}. The continuity equation \eqref{continuity} is in turn a consequence of the Schr\"odinger equation; it is usually written (in standard quantum mechanics) in terms of the quantum probability current $J=\rho \, v$.

\bigskip

\noindent\textbf{Identical Particles.}
Bohmian mechanics can be formulated
for identical particles, despite a fact that could be felt to
contradict their indistinguishability, namely that the particle
trajectories in $\RRR^3$ determine ``who is who'' at different
times, i.e., select a one-to-one association between the $N$
points at any time $t_1$ and the $N$ points at another time $t_2$.
Taking the notion of a particle seriously, as one should in Bohmian mechanics,
one recognizes that the configuration space for $N$ identical particles is
best regarded as the manifold of all sets of $N$ points in physical space $\RRR^3$.
This manifold has non-trivial topological properties, as its
fundamental (homotopy) group is isomorphic to the group
of permutations of $N$ objects. On such manifolds
there arise several versions of Bohmian mechanics corresponding to the
different 1-dimensional representations of the fundamental group; for the permutation group, there are two such
representations, corresponding to bosons (with symmetric wave functions on the covering space $\RRR^{3N}$) and fermions (with anti-symmetric wave functions).
Thus, Bohmian mechanics lends support to the
modern view  that the symmetrization postulate emerges as
a topological effect, due to the non-trivial topology of configuration space.

\bigskip
%\section{Probabilities, Statistics, and Predictions}

\noindent\textbf{Quantum Equilibrium Hypothesis. } This is the assertion that whenever a system has wave function $\psi$ then its configuration is (or can be taken to be) random with probability distribution $|\psi|^2$. Equivariance implies that this hypothesis is consistent with the time evolution of isolated systems, and it is not hard to show that it is also consistent with the time evolution if the system is not isolated, provided we take $\psi$ to mean the conditional wave function. An important consequence of the quantum equilibrium hypothesis is the \emph{empirical equivalence between Bohmian mechanics and quantum mechanics}: For every conceivable experiment, whenever quantum mechanics makes an unambiguous prediction, Bohmian mechanics makes exactly the same prediction. Thus, the two cannot be tested against each other. %For an investigation of how the predictions of Bohmian mechanics might deviate from quantum mechanics if the quantum equilibrium hypothesis was violated, see \cite{Val}.

\bigskip

\noindent\textbf{Typicality. } The quantum equilibrium hypothesis follows from typicality: As shown in \cite{DGZ92} using the law of large numbers, results of experiments are as predicted by the quantum equilibrium hypothesis for \emph{typical} initial configurations $Q(t_0)$ of the universe relative to the $|\Psi_{t_0}|^2$ distribution, i.e., for the overwhelming majority, counted using the $|\Psi_{t_0}|^2$ distribution, of the initial configurations.

\bigskip

\noindent\textbf{Operators. } Given that it makes the same predictions as quantum mechanics, what is the status in Bohmian mechanics of the non-commuting operators of the quantum formalism (the self-adjoint ``observables''), with which the predictions of quantum mechanics seem exclusively concerned? The answer is that operators do in fact arise naturally in Bohmian mechanics, but with a different meaning than the one attributed to them in orthodox quantum mechanics (which regards them as more or less the same thing as their classical counterparts: as ``observables'' that can be ``measured''). Instead, operators in Bohmian mechanics are \emph{mathematical tools encoding statistics}. Let us explain.

The statistics of the random outcome $Z$ of an experiment in a world governed by Bohmian mechanics on a system with wave function $\psi$ can be shown \cite{DGZ04} always to be of the form (in \see Dirac notation)
\begin{equation}
\text{Prob}(Z=\alpha) = \scp{\psi}{E(\alpha)|\psi}\,,
\end{equation}
where $E(\alpha)$ is a suitable positive operator. (Together, the $E(\alpha)$ form a positive-operator-valued measure, or \see POVM.) In relevant cases, $E(\alpha)$ is a family of \see projection operators which are mutually orthogonal (a projection-valued measure, or PVM), and thus correspond to the one self-adjoint operator
\begin{equation}\label{AE}
A = \sum_\alpha \alpha \, E(\alpha)\,,
\end{equation}
which, by the spectral theorem, contains precisely the same information as the PVM $E(\alpha)$.
Thus, operators encode the functional dependence of the outcome statistics on the system's wave function $\psi$. With this understanding, which is opposite to thinking of operators as representing \emph{quantities} whose values can be ``measured,'' it is no longer surprising that one cannot associate \emph{actual values} with all ``observables'' in a consistent way. With this understanding, \emph{contextuality} is not surprising either, since it no longer means that the same quantity can choose different values depending on what happens to another system, but rather that, unspectacularly, different experiments can have the same statistics.

\bigskip

\noindent\textbf{Collapse of the Wave Function.}
Here is an analysis, for Bohmian mechanics, of an ``ideal measurement'' 
of a quantum observable, given by a self-adjoint
operator $A$ on the Hilbert space of the relevant system. For simplicity we assume that $A$ has pure point spectrum with non-degenerate eigenvalues $\alpha$, corresponding to \eqref{AE} for $E(\alpha) = |\psi_\alpha\rangle \langle \psi_\alpha|$ with normalized eigenstates
$\psi_{\alpha}(x)=|A=\alpha\rangle$. The experiment is 
implemented by having the system interact with an apparatus in a
suitable way. To avoid unimportant complications, we shall assume that the relevant ``universe'' for the problem at hand consists entirely of the system, with configuration $X$, and the apparatus, with configuration $Y$.
The measurement begins, say, at time 0, with the initial (``ready'') state of
the apparatus given by a wave function $\varphi_0(y)$, and ends at time
$t$.  The interaction is such that when the state of the system is
initially $\psi_{\alpha}$ it produces a normalized apparatus state
$\varphi_{\alpha}(y)$, that
registers that the value found for $A$ is $\alpha$ without having affected
the state of the system,
\be\label{t}
 \psi_{\alpha}(x)\varphi_0(y)\stackrel{t}{\rightarrow}\psi_{\alpha}(x)\varphi_{\alpha}(y).
\ee
Here $\stackrel{t}{\rightarrow}$ indicates the unitary evolution induced by
the interaction. If the measurement is to provide useful information, the
apparatus states must be noticeably different, corresponding, say,
to a pointer on the apparatus pointing in different directions. We thus
assume that the $\varphi_{\alpha}$ have disjoint supports in the configuration
space for the apparatus,
\be
\text{supp}(\varphi_{\alpha})\cap\text{supp} (\varphi_{\beta})=\emptyset,\  \alpha\neq\beta.
\ee

Now suppose that the system is initially, not in an eigenstate of $A$, but in
a general state, given by a superposition
\be\label{11}
\psi(x)=\sum_{\alpha}c_{\alpha}\psi_{\alpha}(x).
\ee
We then have, by \eqref{t} and the linearity of the unitary evolution, that
\be
\Psi_0(x,y)=\psi(x)\varphi_0(y)\stackrel{t}{\rightarrow}\Psi_t(x,y)=\sum_{\alpha}c_{\alpha}\psi_{\alpha}(x)\varphi_{\alpha}(y),
\ee
so that the final wave function $\Psi_t$ of system and apparatus is itself
a superposition. The fact that the pointer ends up pointing in a definite
direction, even a random one, is not discernible in this final wave
function. Insofar as orthodox quantum theory is concerned, we have arrived at
the measurement problem.

However, insofar as Bohmian mechanics is concerned, we have no such
problem, because in Bohmian mechanics particles always have positions and
pointers, which are made of particles, always point---in a direction
determined by the final configuration $Y_t$ of the apparatus. Moreover, in
Bohmian mechanics we find that the state of the system is transformed in
exactly the manner prescribed by textbook quantum theory, as 
the final wave function of the system, i.e., its conditional wave function at time $t$, see \eqref{condwf}, is
\be\label{13}
\psi_t(x)=%\mathcal{N}\,
\Psi_t(x,Y_t)=%\mathcal{N}
\sum_{\alpha}c_{\alpha}\psi_{\alpha}(x)\varphi_{\alpha}(Y_t)=%\mathcal{N}\,
c_\beta\psi_\beta(x)\varphi_\beta(Y_t)
=\mathcal{N}\, \psi_\beta(x)
\ee
when $Y_t\in\text{supp}(\varphi_\beta)$, 
i.e., when the value $\beta$ is registered. 
(Here $\mathcal{N}$ is a constant that depends upon $Y$ but not on $x$. According to \eqref{13} the wave function of the system at time $t$, when normalized, is $\psi_\beta$.) 
The probability for this event is, by the quantum equilibrium hypothesis,
\be
\int dx\int\limits_{\text{supp}(\varphi_\beta)} dy\,|\Psi_t(x,y)|^2 = |c_\beta|^2.
\ee
The upshot of the analysis is this: It is a consequence of Bohmian mechanics that in the course of an ideal measurement of $A$ the (normalized) wave function of the system is transformed from $\psi$ \eqref{11} to $\psi_\beta$ with probability $|c_\beta|^2 = \bigl| \scp{\psi_\beta}{\psi} \bigr|^2$. That is how the \see \emph{projection postulate} arises from Bohmian mechanics. (The fact that the contributions with $\alpha\neq \beta$ will never again overlap with what evolves from $\psi_\beta(x)\varphi_\beta(y)$, and thus will not influence the future motion of the particles, is the reason why they can be ignored from time $t$ onwards, or ``collapsed away,'' without consequences for the trajectories of the particles.)

\bigskip
%\section{Quantum Phenomena in the Bohmian Way}

\noindent\textbf{The Double Slit Experiment.}
In Bohmian mechanics, ``wave--particle duality'' can be taken literally: there is a wave ($\psi$) and there are particles. Accordingly, in a double slit experiment the wave passes through both slits, whereas the particle passes only through one slit. Since the motion of the particle depends on the wave, it matters whether or not the other slit is open. The possible trajectories, when both slits are open, are depicted in Fig.~1; by virtue of the quantum equilibrium hypothesis, the actual trajectory will be random with the appropriate $|\psi|^2$ distribution. Thus, the place of the particle's arrival at a screen on the right will have a probability distribution featuring interference fringes. As John Bell commented \cite[p.~191]{Bell87b}: ``This idea seems to me so natural and simple [...] that it is a great mystery to me that it was so generally ignored.''

\begin{figure}[ht]\begin{center}\includegraphics[width=.5 \textwidth]
{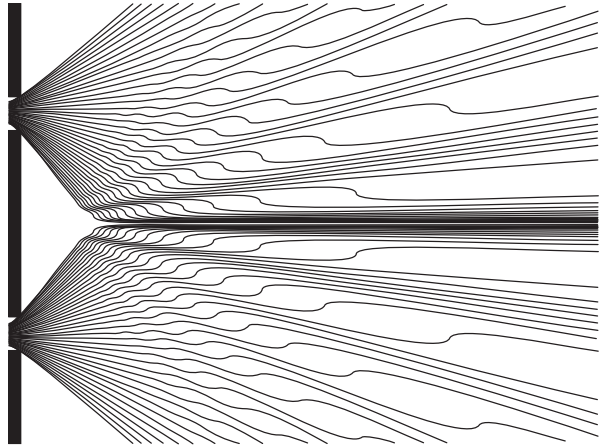}\end{center}\caption{Possible Bohmian trajectories in the double-slit experiment (from C.\ Philippidis, C.\ Dewdney and B.J.\ Hiley, {\em Il Nuovo Cimento}
{\bf 52}, 15 (1979))}\end{figure}

\bigskip

\noindent\textbf{Spin.}
One may easily get the impression that spin cannot be explained in a
realist way, given its ``non-classical two-valuedness.'' But actually it can be incorporated into Bohmian mechanics very easily, and Bell discovered how \cite{Bell66}: Do not assume that there is an ``actual value'' associated with the spin observable $\hat\sigma_z$ in the $z$ (or any other) direction!  Instead, take the equation of motion \eqref{Bohm} seriously, with $\CCC^k$ the spin space, i.e., $k=(2s+1)^N$ for $N$ spin-$s$ particles. (In particular, it is useful here to regard the wave function $\psi_t$ for, say, a single spin-$\tfrac{1}{2}$ particle not as a function $\psi_t:\RRR^3\times \{-1,1\} \to \CCC$ 
of a continuous (position) variable and a discrete (spin) variable, but rather as a spinor-valued function of position, $\psi_t: \RRR^3\to\CCC^2$.)

As a consequence of \eqref{Bohm}, the motion of a particle with spin is influenced by both the ``spin-up'' and the ``spin-down'' component of the wave function. While the particle has an actual position (and a wave function) but no additional actual spin degrees of freedom, these are sufficient to completely account for all quantum  phenomena associated with spin. %While there exists no direction in which the spin ``actually'' points, the particle always has a position, and that is enough for the purpose of obtaining a clear theory.

\bigskip

\noindent\textbf{Quantum Field Theory and Relativity.}
Bohmian mechanics does not account for phenomena such as particle  
creation and annihilation characteristic of quantum field theory.  
This is not an objection to Bohmian mechanics but merely a  
recognition that quantum field theory explains a great deal more than  
does nonrelativistic quantum mechanics, whether in orthodox or  
Bohmian form. There are extensions  of Bohmian mechanics to general  
quantum field theories based on a particle ontology, as well as
other approaches.
Moreover, like nonrelativistic quantum theory, Bohmian mechanics is  
incompatible with special relativity, a central principle of physics:  
it is not Lorentz invariant. Nor can Bohmian mechanics easily be  
modified to become Lorentz invariant. For an overview of recent proposals aimed at  
finding a Lorentz invariant extension of Bohmian mechanics, see \cite{Tum06b}.

\bigskip

\noindent\textbf{Nonlocality.} 
In Bohmian mechanics the motion of a particle may depend on the positions of distant particles, at spacelike separation. This is an instance of \emph{nonlocality}. It is worth noting that this dependence is of a kind that does not allow superluminal communication. Orthodox quantum mechanics features nonlocality as well, associated with the instantaneous collapse of the wave function for all particles, even distant ones. In 1964, John Bell asked whether nonlocality could be avoided by any version of quantum mechanics, and his celebrated (but often misunderstood) argument \cite{Bell64,Bell87b}, involving \emph{Bell's inequality}, proves that the answer is no. His argument shows that certain correlations predicted by quantum mechanics (and Bohmian mechanics) and confirmed in experiment \cite{Aspect1982} cannot be explained in a local way, i.e., without allowing influences travelling faster than light. Thus, nonlocality is a feature of our world.%, and makes the existence of a time foliation seem a serious possibility. 


\begin{thebibliography}{100}

\item[\it Primary]

\bibitem{Aspect1982} A.~Aspect, J.~Dalibard, G.~Roger:
    Experimental Test of Bell's Inequalities using Time-Varying
    Analyzers. \textit{Physical Review  Letters} \textbf{49}, 1804--1807
    (1982)

\bibitem{Bell66} J. S. Bell: On the problem of hidden variables in
  quantum mechanics. \textit{Reviews of Modern Physics} \textbf{38}, 447--452
  (1966); reprinted as chapter 1 of \cite{Bell87b}.

\bibitem{Bell64} J.~S. Bell: On the
    Einstein-Podolsky-Rosen Paradox.  \textit{Physics},
    \textbf{1}, 195--200 (1964); reprinted as chapter 2 of \cite{Bell87b}.
    
\bibitem{Bell86} J. S. Bell: Quantum field theory without observers.
  \textit{Physics Reports} \textbf{137}, 49--54 (1986); reprinted 
  under the title ``Beables for quantum field theory'' as
  chapter 19 of \cite{Bell87b}.

%\bibitem{Bell86b} Bell, J. S.:
%  Six possible worlds of quantum mechanics.
%  In \textit{Proceedings of the Nobel Symposium 65: Possible Worlds in Arts and Sciences} (Stockholm, August 11--15, 1986). 
%  Reprinted as chapter 20 in \cite{Bell87b}.%, p.~181.

%\bibitem{Bell87} Bell, J. S.: Are there quantum jumps? In
%  \textit{Schr\"odinger. Centenary Celebration of a Polymath.} Cambridge:
%  Cambridge University Press (1987). Reprinted as chapter 22 of
%  \cite{Bell87b}.

\bibitem{Bohm52} D. Bohm: A Suggested Interpretation of the Quantum
  Theory in Terms of ``Hidden'' Variables, I and II. \textit{Physical
    Review} \textbf{85}, 166--193 (1952)

%\bibitem{BHK} D. Bohm, B. J. Hiley, P. N. Kaloyerou:
%   An ontological basis for the quantum theory. 
%   \textit{Physics Reports} \textbf{144},  321--375 (1987)

%\bibitem{CS07} S. Colin, W. Struyve: 
%	 A Dirac sea pilot-wave model for quantum field theory.
%	\textit{Journal of Physics A: Mathematical and Theoretical} 
%	\textbf{40}, 7309--7341 (2007) 
%	arXiv:quant-ph/0701085

\bibitem{deB} L. de Broglie: 
	In %Solvay Congress (1927),
  {\em Electrons et Photons: Rapports et Discussions du Cinqui\`eme
    Conseil de Physique tenu \`a Bruxelles du 24 au 29 Octobre 1927
    sous les Auspices de l'Institut International de Physique Solvay},
  (Gauthier-Villars, Paris 1928); English translation in \cite{BV}.
  
%\bibitem{HBD} D. D\"urr, S. Goldstein, K. M\"unch-Berndl, N. Zangh\`\i: 
%  Hypersurface Bohm--Dirac Models. 
%  \textit{Phys. Rev. A} \textbf{60}, 2729--2736 (1999). %arXiv:quant-ph/9801070

%\bibitem{DGTTZ07}
%    D. D\"urr, S. Goldstein, J. Taylor, R. Tumulka, N. Zangh\`\i:
%    Quantum Mechanics in Multiply-Connected Spaces.  
%    \textit{Journal of Physics A: Mathematical and Theoretical}
%    \textbf{40}, 2997--3031 (2007) %arXiv:quant-ph/0506173.

%\bibitem{DGTZ05} D. D{\"u}rr, S. Goldstein, R. Tumulka, N.
%  Zangh{\`{\i}}: Bell-Type Quantum Field Theories.
%  \textit{Journal of Physics A: Mathematical and General} \textbf{38}, R1--R43 (2005)
  %arXiv:quant-ph/0407116

\bibitem{DGZ92} 
  D. D\"urr, S. Goldstein, N. Zangh\`\i:
  Quantum Equilibrium and the Origin of Absolute Uncertainty. 
  \textit{Journal of Statistical Physics} \textbf{67}, 843--907 (1992) 
  %arXiv:quant-ph/0308039 

\bibitem%[D\"urr et al.(2004)D\"urr, Goldstein, and Zangh\`\i{}]
{DGZ04}
  D. D\"urr, S. Goldstein, N. Zangh\`\i: Quantum
  Equilibrium and the Role of Operators as Observables in Quantum 
  Theory. \textit{Journal of Statistical Physics} \textbf{116}, 959--1055 (2004)
  %quant-ph/0308038

%\bibitem{LM77}
%J.~Leinaas, J.~Myrheim: 
%On the theory of identical particles. 
%\textit{Il Nuovo Cimento} \textbf{37 B}, 1--23 (1977) 

%\bibitem{Rosen} N. Rosen: On Waves and Particles. 
%  \textit{Journal of the Elisha Mitchell Scientific Society} \textbf{61}, 67--73 (1945)

%\bibitem{SW06} W. Struyve, H. Westman:
%  A New Pilot-Wave Model for Quantum Field Theory. 
%  In A. Bassi, D. D\"urr, T. Weber and N. Zangh{\`{\i}} (eds.), 
%  \textit{Quantum Mechanics: Are there Quantum Jumps? and
%  On the Present Status of Quantum Mechanics}, 
%  AIP Conference Proceedings \textbf{844} (American Institute of Physics
%  2006, 321--339) %quant-ph/0602229

%\bibitem{Val} A. Valentini: Subquantum Information and Computation.
%  {\em Pramana - Journal of Physics} \textbf{59}, 269--277 (2002)


\item[\textit{Secondary}]



\bibitem{BV} G. Bacciagaluppi, A. Valentini: 
  \textit{Quantum Theory at the Crossroads} (Cambridge 
  University Press 2009)



\bibitem{Bell87b} J. S. Bell: \textit{Speakable and unspeakable in
    quantum mechanics} (Cambridge University Press 1987)

\bibitem{BH93} D. Bohm, B.J. Hiley: 
  \textit{The Undivided Universe} 
  (Routledge, London 1993)  

\bibitem{Gol01} S. Goldstein: Bohmian Mechanics. 
  In E. N. Zalta (ed.), \textit{Stanford Encyclopedia of Philosophy}, 
  published online by Stanford University (2001) at\\
  http://plato.stanford.edu/entries/qm-bohm/

%\bibitem{DL71}
%  M.~G. Laidlaw, C.~M. DeWitt: 
%  Feynman functional integrals for systems of indistinguishable particles. 
%  \textit{Physical Review D} \textbf{3}, 1375--1378 (1971) 

%\bibitem{Sch81}
%L.~S. Schulman: 
%\textit{Techniques and Applications of Path Integration}. 
%(John Wiley \& sons, New York 1981)

%\bibitem{Struyve07} W. Struyve:  Field beables for quantum field theory.\\
%	 http://arXiv.org/abs/0707.3685

\bibitem{Tum06b} R. Tumulka:
	The `Unromantic Pictures' of Quantum Theory.
	\textit{Journal of Physics A: Mathematical and Theoretical} 
	\textbf{40}, 3245--3273 (2007)
%	arXiv:quant-ph/0607124

\end{thebibliography}
\end{document}